\documentstyle[times, xspace,graphicx]{jaa}
\providecommand{\xray  }{X-ray\xspace}%
\providecommand{\xrays }{X-rays\xspace}%
\providecommand{\gray  }{$\gamma$-ray\xspace}%
\newcommand\apj{\textrm{ApJ}}%
\newcommand\aap{\textrm{A\&A}}%
\newcommand\mnras{\textrm{MNRAS}}%
\begin{document}
\title[\xray Time Lags in TeV Blazars]{\xray Time Lags in TeV Blazars} 
\author[X. Chen et al.]%
       {X.~Chen$^1$\thanks{e-mail:xuhui@rice.edu},
        G.~Fossati$^1$, 
        E.~Liang$^1$
        and M.~B\"ottcher$^2$ \\ 
$^1$ Department of Physics and Astronomy, Rice University, Houston, Texas, USA 77005 \\
$^2$ Astrophysical Institute, Department of Physics and Astronomy, Ohio University
}
\maketitle
\label{firstpage}
\begin{abstract}
We use Monte Carlo/Fokker-Planck simulations to study the \xray time lags. 
Our results show that soft lags will be observed as long as the decay of the
flare is dominated by radiative cooling, even when acceleration and cooling
timescales are similar. 
Hard lags can be produced in presence of a competitive achromatic particle
energy loss mechanism if the acceleration process operates on a timescale such
that particles are slowly moved towards higher energy while the flare evolves.
In this type of scenario, the \gray/\xray quadratic relation is also reproduced.

\end{abstract}


\section{Introduction}
\label{sec:intro}

One of the most interesting and least studied in details aspects of TeV blazars
variability is that of time lags between variations at different energies in
the \xray band.
The results of the time delay analysis of the \xray
sub-bands include all three possibilities: soft lag, hard lag, and no lag. (e.g.
Rebillot et al., 2006;
Fossati et al., 2000;
Brinkmann et al, 2005).
\nocite{rebillot_etal:2006:multiwavelength_mrk421,%
fossati_etal:2000:mkn421_temporal,%
brinkmann_etal:2005:mrk421_xmm}

One the best modeling works addressing the issue of time lags in the
synchrotron emission remains that by Kirk and collaborators (1998).
\nocite{kirk_rieger_mastichiadis:1998}
They assumed that most of the jet emission, certainly the more highly variable
component, is produced by shocked plasma and modeled it as a moving shock and
its immediate downstream region.
Their work, similarly to most more recent modeling, did not take into account
the light travel times effects (LTTE) within the emission region and with
respect to the observer, which can modify significantly the phenomenology 
\nocite{chen_etal:2010:multizone_code_mrk421} (e.g., Chen et al., 2010).
LTTE are important to blazars but very difficult to calculate in
traditional ways of solving the radiative transfer equation.

\begin{figure}[t]
\centerline{%
\hfill
\includegraphics[width=0.52\linewidth]{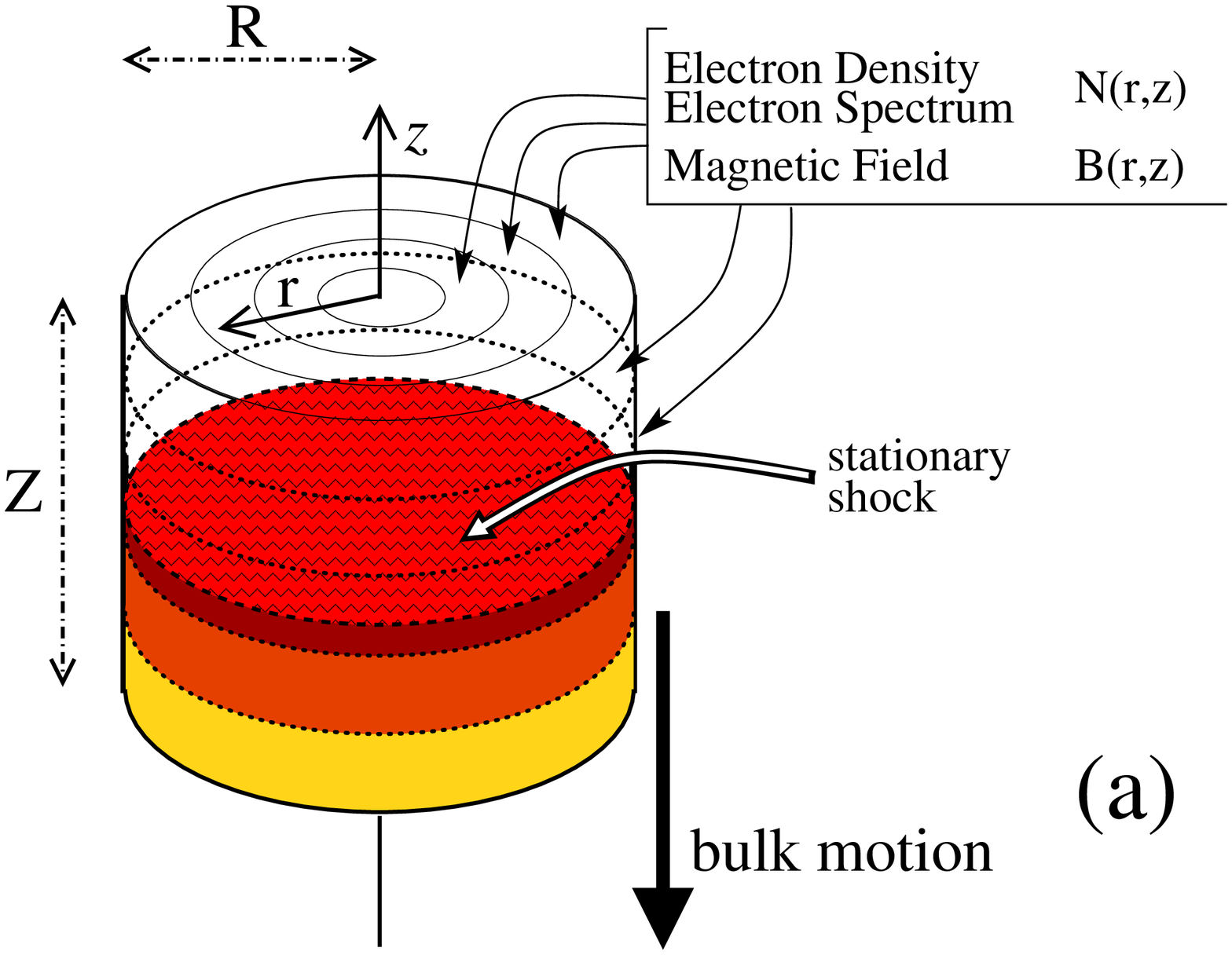}
\hfill
\includegraphics[width=0.42\linewidth]{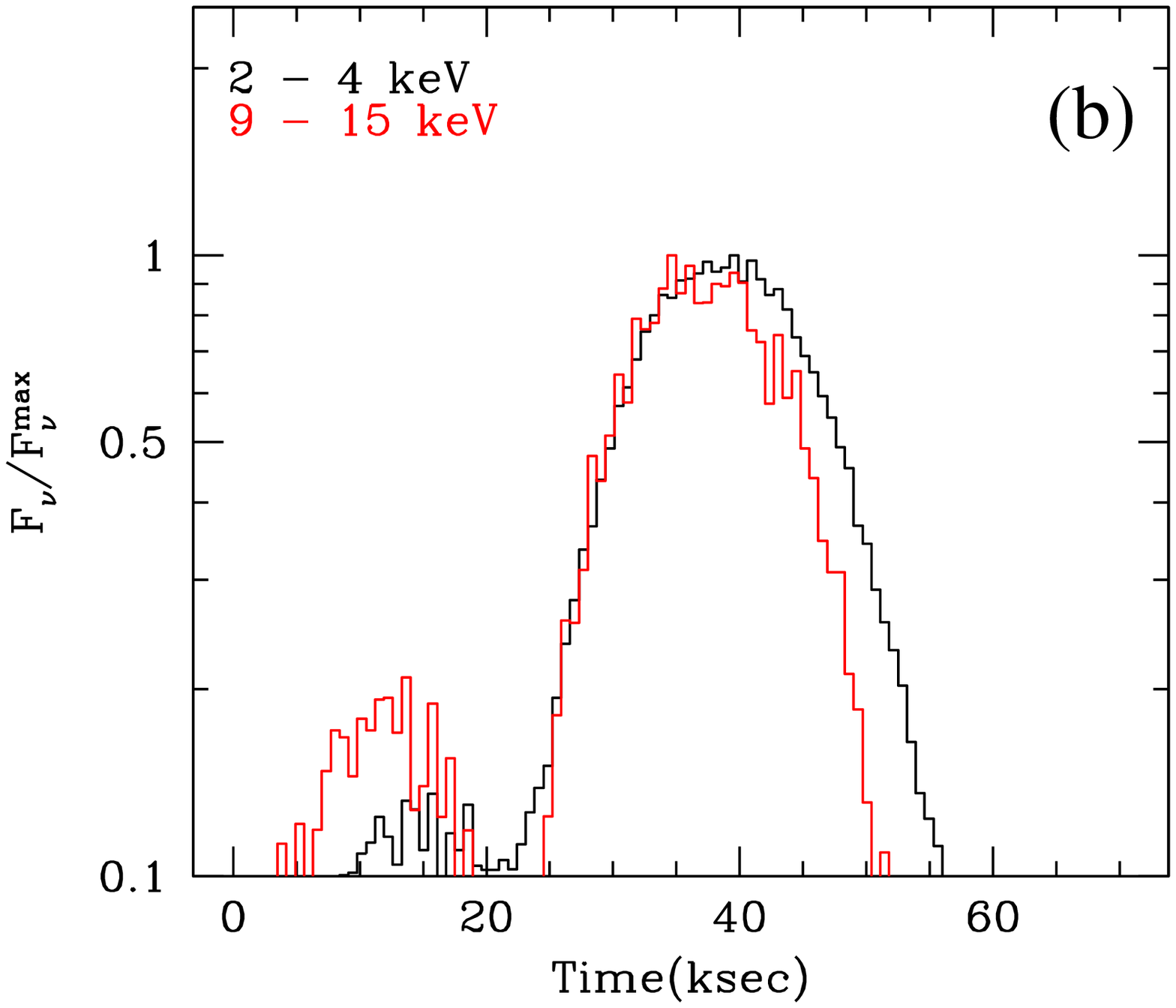}
\hfill
}
\vspace{-4mm}
\caption{%
(a): The geometry of the model. The volume is divided in the $r$ and
$z$ directions in zones with their own electron distribution and magnetic field.
We also schematically show the setup for the variability of the simulations
with a shock.
The hatched layer represents a stationary shock.
The blob moves downward and crosses the shock front.
Zones that crossed the shock at earlier times have had some time to radiate
the newly injected energy and are plotted in lighter color shades. 
(b): The light curves produced by the scenario \#2. During the flare rise both
bands vary together, but in the decay phase the harder band drops more rapidly
yielding a soft lag.
\label{fig:geometry_and_lc}
}
\end{figure}

Here we present preliminary results of a more advanced simulation code that
allows us to include LTTE, and begin to analyze the scenarios and conditions
that can lead to different time-lag signatures.
%
The results are obtained with our Monte Carlo/Fokker-Planck radiation transfer
code described in \nocite{chen_etal:2010:multizone_code_mrk421} (Chen et al.,
2010).
The Monte Carlo methods allows to track the trajectories of individual
(pseudo)photons, thus accounting naturally for all the LTTEs.
We handle the electrons as populations with densities and distributions, and
use the Fokker-Planck equation to solve for their time evolution. 
The acceleration, cooling, injection and achromatic loss of electrons are all
realized through the Fokker-Planck equation.
The geometry of the plasma blob is assumed to be cylindrical as shown in
Fig.~\ref{fig:geometry_and_lc}, left.

\begin{figure}[t]
\centerline{
\hfill
\includegraphics[width=0.49\linewidth]{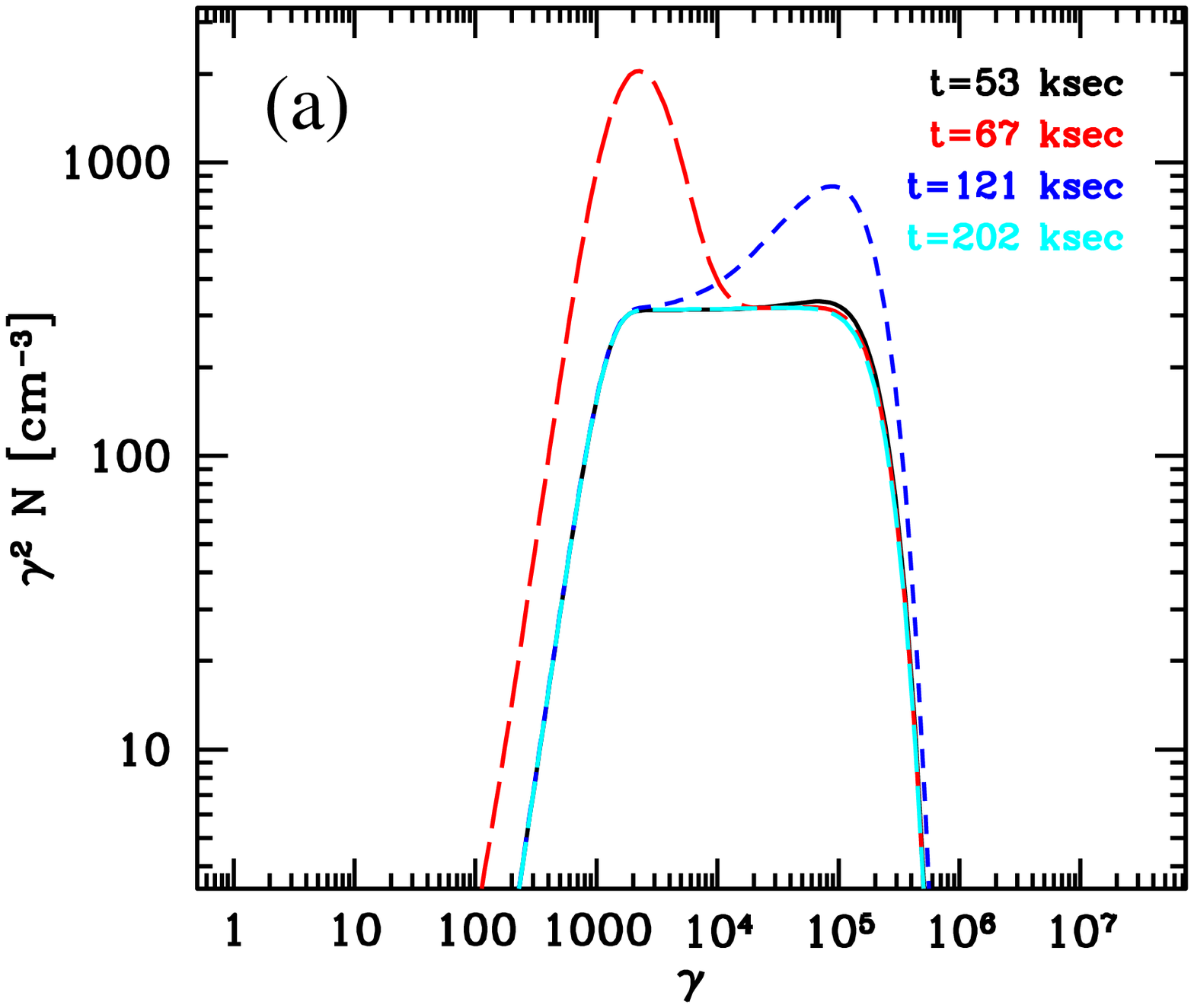}
\hfill
\includegraphics[width=0.49\linewidth]{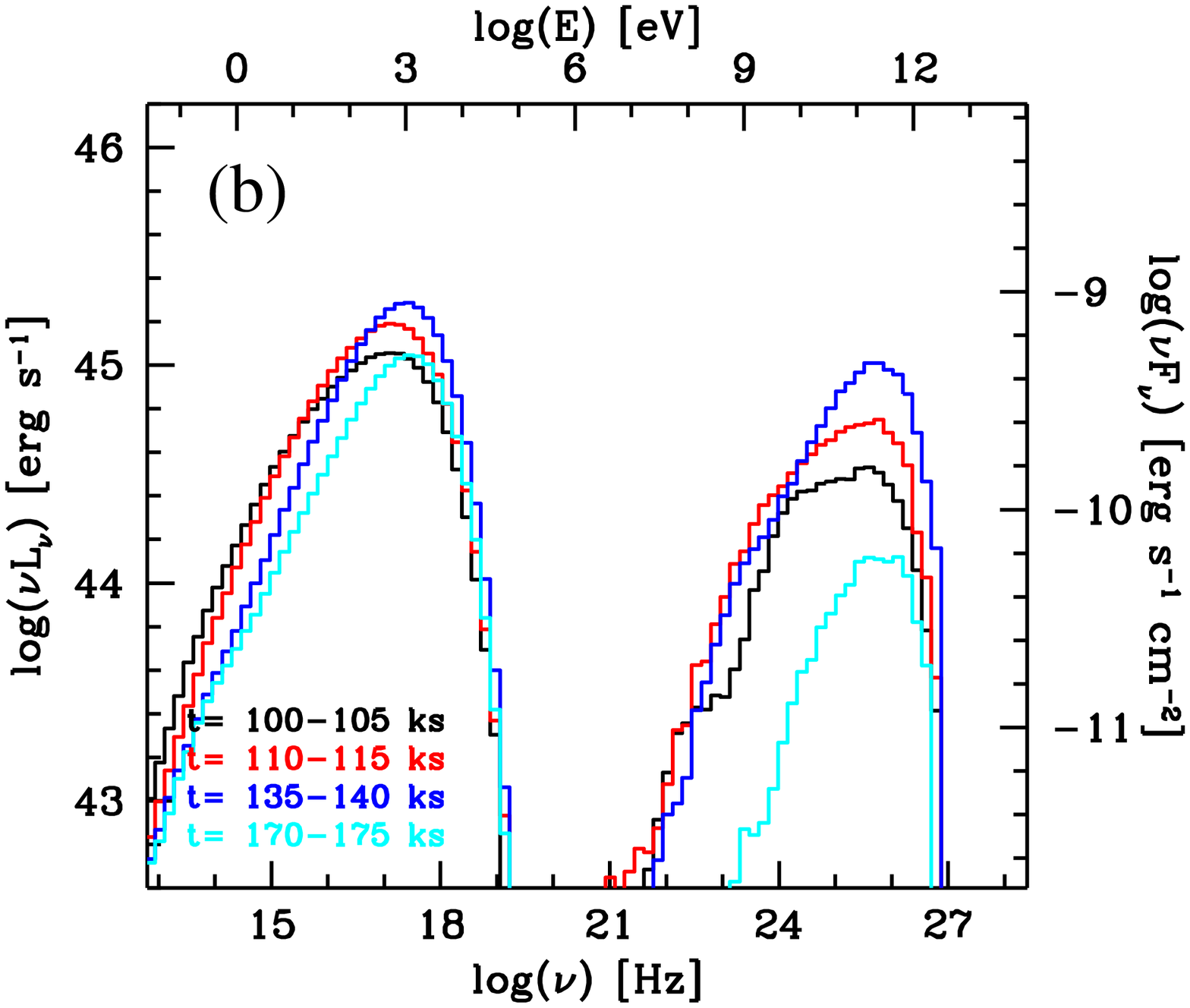} 
\hfill
}
\vspace{-4mm}
\caption{
For case \#\,4. 
(a): evolution of the electron spectrum in one of the blob zones.
(b): SEDs.
In both panels, the time sequence is: black, red, blue, cyan.
Times are in the observer's frame.
Model parameters: 
$B=0.1$\,G, 
$\delta=33$,
sizes: $z=1.33\times10^{16}$\,cm, $r=10^{16}$\,cm, 
$n_e=0.4$\,cm$^{-3}$,
injection rate $q=3.17\times10^{39}$\,erg/s without shock 
increasing by a factor of 8 with the shock.
Time-scales: acceleration $t_{\rm acc}=z/c$, and
achromatic loss $t_{\rm loss}=z/c$. 
The shock begins to cross the blob at $t=60$\,ks.
\label{fig:case4_electrons_and_seds}
}
\end{figure}

\section{Summary of four flare scenarios}
\label{sec:results}

We have tested 4 different scenarios:
  
\noindent
\textbf{\#1}: Homogeneous, steady rate, injection of high energy particles
with power law distribution.
The flare is caused by an increase/decrease of the maximum electron energy
$\gamma_{\rm max}$, following an exponential (symmetric) time evolution.
 
\noindent
\textbf{\#2}: Homogeneous {\it mild} diffusive particle acceleration mechanism
is active in the blob for a set duration, after which the evolution is purely
radiative.

\noindent
\textbf{\#3}: Rapid electron acceleration locally as the shock crosses the blob,
followed by a purely radiative evolution.

\noindent
\textbf{\#4}: The shock causes a local burst of injection of {\it medium}
energy electrons ($\gamma = 10^3$, with narrow Gaussian spectrum), which
happens in a blob where it a steady diffusive (slow-ish) acceleration mechanism
is present.
The particle cooling is not purely radiative, but it includes an achromatic
energy loss process which is the main factor controlling the flare decay.

\smallskip
The physical explanation for the {\it mild} diffuse acceleration in the blob can
be shear acceleration \nocite{rieger_duffy:2004:shear_acceleration} 
(e.g., Rieger \& Duffy, 2004), while the injected {\it medium} energy electrons
may come from the stochastic particle acceleration (e.g., Katarzy\'nski et al. 2006).
\nocite{katarzynski_etal:2006:stochastic}
The achromatic energy loss can be thought of as caused by adiabatic expansion
or particle escape.

The first three models failed to produce a \xray hard lag (e.g., see the light
curves for \#2 in Fig.~\ref{fig:geometry_and_lc}b). 
In cases \#1 and \#2 the soft \xray variation leads the hard \xray one, but
in all these models the spectral evolution during the flare decay is controlled
by radiative cooling, hence it always propagates from high to low energy,
yielding a soft lag (even after smearing by LTTEs).

\begin{figure}[t]
\centerline{
\hfill
\includegraphics[width=0.46\linewidth,clip=]{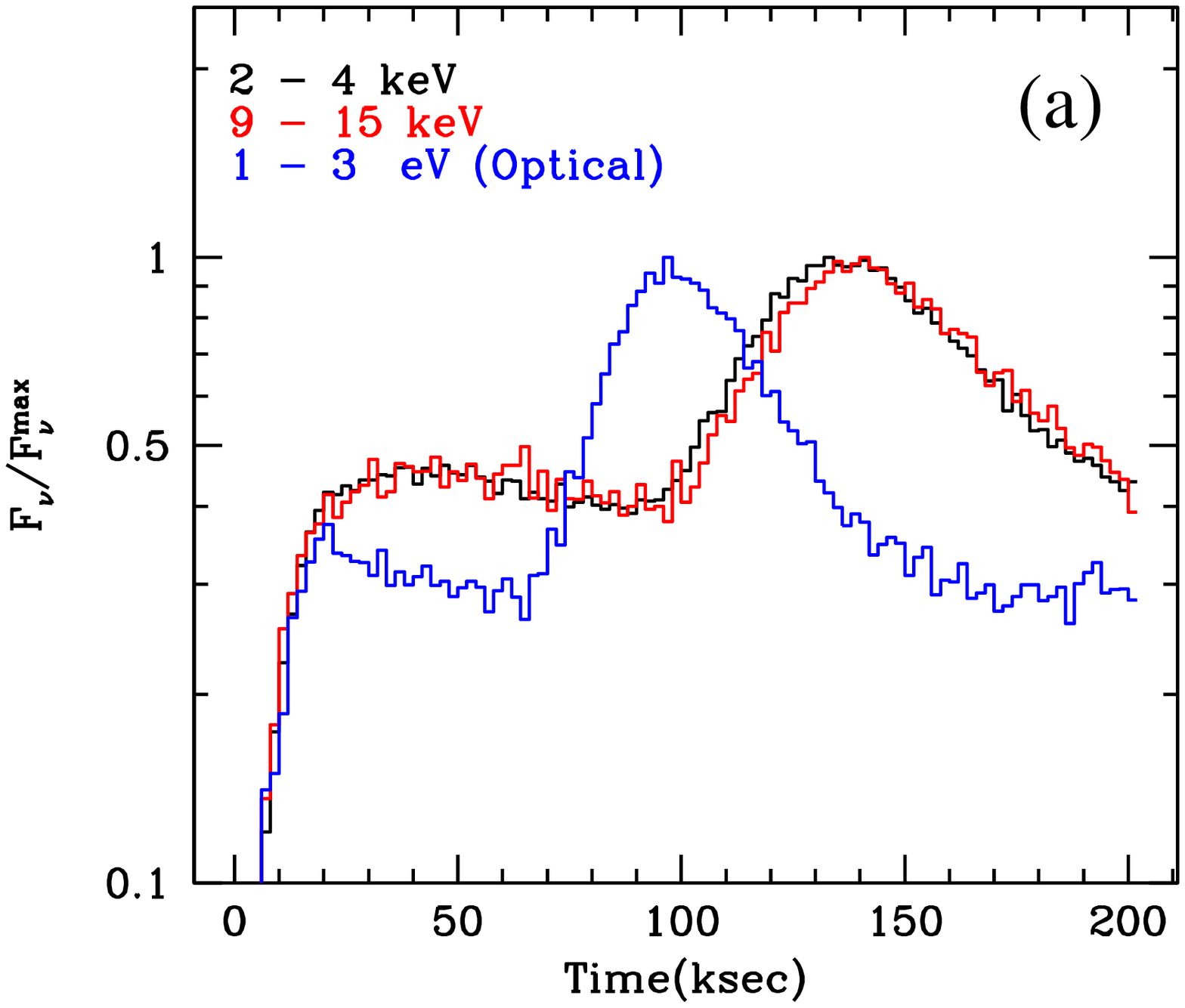}
\hfill
\includegraphics[width=0.46\linewidth,clip=]{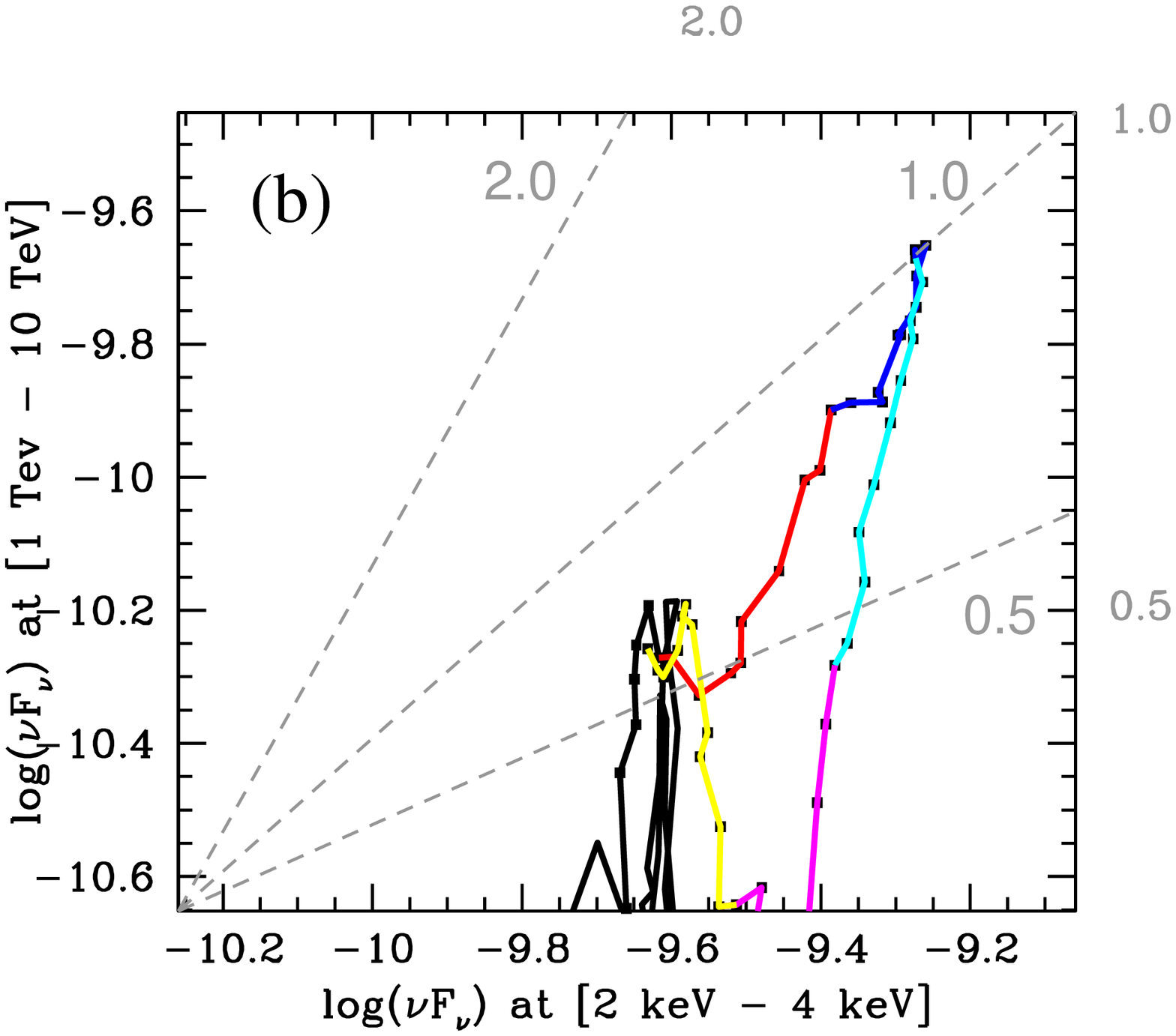}
\hfill
}
\caption{
Case \#4. 
(a): Light curves in three energy bands. 
(b): \gray vs. \xray flux correlation.
Colors represent different times: 
black ($<100ks$), 
red ($100-120$ ks),
blue ($120-140$ ks), 
cyan ($140-160$ ks), 
magenta ($160-180$ ks), 
yellow ($>180$ ks).
\label{fig:case4_lc_and_ff}
}
\end{figure}

The more complex fourth scenario successfully produced a hard \xray lag 
(see Fig.~\ref{fig:case4_electrons_and_seds}b for SEDs and
Fig.~\ref{fig:case4_lc_and_ff}a for light curves).
This model is similar to the second model, in the sense that in both the {\it
mild} particle acceleration slowly moves the electrons from low/medium energy
to high energy, providing the hard lag when the flux increases.
The crucial difference is that the decay of the flare in the fourth model is
controlled by achromatic energy loss, which eliminates the emergence of the
soft lag.
Radiative cooling is balanced by particle acceleration, so it does not
have much control on the flux and spectral change.
The left panel of Fig.~\ref{fig:case4_lc_and_ff} shows how the electron
energy distribution evolves in this model.

A potential problem in this model is that the optical flux shows a large, and
early, variation, as seen in Fig. \ref{fig:case4_lc_and_ff}a, which is not
usually observed.
This problem can be mitigated by the possible presence of additional emission
by other regions of the jet having lower energy particles (e.g. Ushio et al.,
2009; Krawczynski, Coppi \& Aharonian, 2002; Chen et al., 2010).
\nocite{%
krawczynski_coppi_aharonian:2002:timedep,%
ushio_etal:2009:mrk421_in_2006_with_suzaku,%
chen_etal:2010:multizone_code_mrk421}
This emission can have a SED peaking closer to the optical band and not very
luminous in \xrays, thus diluting significantly the observed optical variations
with respect to their intrinsic magnitude, without affecting much our view of
the flaring emission in \xray.

Scenario \#4 also yields a quadratic relation between \gray and \xray fluxes in
both the raising and decay phases of the flare (see Fig.~\ref{fig:case4_lc_and_ff}b), 
a feature frequently observed in TeV blazars that has proved to be challenging
to model (Fossati et al., 2008). \nocite{fossati_etal:2008:xray_tev} 
Our previous efforts with flare evolution dominated by radiative cooling could
only produce the quadratic relation for the flare rising phase (Chen et al., 2010). 
\nocite{chen_etal:2010:multizone_code_mrk421} 
Again the crucial element here is the adiabatic energy loss mechanism, because
it affects at the same time the medium energy electrons that emit the seed
photons and the high energy electrons that inverse Compton scatter them.


\def\newblock{\hskip .11em plus .33em minus .07em}


\label{lastpage}
\end{document}